\documentclass[preprint]{revtex4}
\begin{document}
\title{A solution of the cosmological constant problem}
\date{\today}
\author{E. F. Ferrari}
\affiliation{Departamento de F\'{\i}sica, Universidade Federal do
Paran\' a, C.P. 19081, 81531-990, Curitiba, PR, Brazil}
\begin{abstract}
General relativity can be cast as a gauge theory by introducing a tetrad
field and a spin-connection. This formalism was extended by replacing
the tetrad field with a mixed tensor field independent of the metric tensor
in order to develop a mechanism of adjustment of the vacuum energy density
that takes advantage of Weinberg's no-go theorem.
With no anthropic considerations, it was shown that the vacuum energy
density is bounded and the gravitational and cosmological constants are
proportional to a tiny dimensionless parameter determined by the coupling
constants of the model.
\end{abstract}
\maketitle
Recent observations indicate that in the Planck scale the effective
cosmological constant is $8\pi G\,\Lambda\approx10^{-120}$.
The problem of the cosmological constant is to understand why it is so
small.\cite{Weinberg89,Weinberg96,Rugh,Carroll,Garriga,Straumann,%
Peebles,Pad}
Since any contribution to the energy density in the vacuum acts like
a cosmological constant, quantum field theory leads us to expect an
enormously larger value.
In the standard model of particle physics, the sum of the zero-point
energies of the normal modes of the fields up to a wave number cutoff of
100 GeV makes a contribution in excess of fifty-six orders of magnitude.
Setting Einstein's cosmological constant to cancel this contribution
requires an extreme fine-tuning that appears unnatural.
Although several tentative solutions have been proposed, the tiny value
of the effective cosmological constant remains a deep mystery of
fundamental physics.

An adjustment by spontaneous symmetry-breaking seems the most promising
solution, ``an idea that has been tried by virtually everyone who
has worried about the cosmological constant''.\cite{Weinberg89}
(A long list of references can be found in \cite{Dolgov}.)
Weinberg's no-go theorem should, however, get due attention:
a solution for one or more scalar fields does not imply the vanishing of
$\Lambda$.\cite{Weinberg89,Weinberg96}
From the point of view adopted in this paper, this should be considered
a positive result. After all, the effective cosmological constant accounts
for 0.7 of the present energy density of the universe.
So, instead of trying to get rid of $\Lambda$, let us assume that it is
simply related to the vacuum expectation value of a scalar field,
$\Lambda=\gamma\phi_{vac}^2$.
In the tetrad formalism (see below), the Einstein field equation for pure
gravity with this cosmological constant takes the form
$R^{ab}_{~~\mu\nu}\,e^{\nu}_b=-\gamma\phi_{vac}^2\,e^a_{\mu}$,
with no explicit contribution of the energy density of matter
in the vacuum.
The simplest way of obtaining this equation is by assuming that
it is valid in general, but, of course, this assumption conflicts with
the Einstein field equation including matter.
To avoid this inconsistency, we have to introduce in the place of the
tetrad a mixed tensor field $V^a_\mu$ independent of the metric tensor
and the spin-connection. The adopted Lagrangians describe the simplest
model in which this idea can be implemented.

In natural units, the cosmological constant has the dimension of
$[\text{energy}]^2$ and the gravitational constant has the dimension of
$[\text{energy}]^{-2}$.
As $\Lambda$ and $8\pi G$ have reciprocal dimensions,
it is possible to define a unit of energy such that they have the same
numerical value, $\Lambda=\text{tiny number}~[\text{unit of energy}]^2$
and $8\pi G=\text{same tiny number}~[\text{unit of energy}]^{-2}$,
just choose the unit of energy such that
$\Lambda / 8\pi G = 1~[\text{unit of energy}]^4$. 
What is the physics determining this energy scale and this tiny number?
As we shall see below, the energy scale is determined by the maximum of
the potential energy of the scalar field that spontaneously breaks
the symmetry of the vacuum and the tiny number is given by
the dimensionless coupling constants of the model.
This result is not substantially affected by the energy density of
matter in the vacuum, whose range of possible values is bounded. 
In this sense, the proposed mechanism of adjustment of the total energy
density in the vacuum is a solution of the cosmological constant
problem.

As everyone knows, general relativity can be cast as a gauge theory
by introducing a tetrad field $e^a_\mu$ with the properties
\begin{eqnarray}
\label{eq:tetrad1}
g_{\mu\nu}=\eta_{ab}e^a_\mu e^b_\nu~,\\
\label{eq:tetrad2}
\eta^{ab}=g^{\mu\nu}e^a_\mu e^b_\nu~,
\end{eqnarray}
where $\eta_{ab}=\eta^{ab}=\text{diag}(1,-1,-1,-1)$.
These equations are invariant under local $SO(1,3)$ transformations of
the tetrad, all corresponding to the same space-time metric.
The spin-connection $\omega^a_{~b\mu}$ is the gauge potential of
this symmetry and the gauge field strength is 
\begin{equation}
R^a_{~b\mu\nu}=\partial_\mu\omega^a_{~b\nu}
-\partial_\nu\omega^a_{~b\mu}
+\omega^a_{~c\mu}\omega^c_{~b\nu}
-\omega^a_{~c\nu}\omega^c_{~b\mu}~.
\end{equation}
The Hilbert-Palatini Lagrangian density expressed in terms of the tetrad
and the spin-connection is
\begin{equation}
\label{eq:HP}
\mathcal{L}_{HP}=-\frac{1}{16\pi G}\,e\,e^\mu_ae^\nu_b
R^{ab}_{~~\mu\nu}~,
\end{equation}
where $e=\det e^a_\mu$, $e^\mu_a=g^{\mu\nu}\eta_{ab}e^b_\nu$, and
$R^{ab}_{~~\mu\nu}=\eta^{bc}R^a_{~c\mu\nu}$.
From the corresponding action for pure gravity with
a cosmological constant, the Einstein field equation takes the
form\cite{Peldan}
\begin{equation}
\label{eq:einstein}
R^{ab}_{~~\mu\nu}\,e^{\nu}_b=-\Lambda\,e^a_{\mu}~,
\end{equation}
where $\omega^a_{~b\mu}$ is the unique torsion-free spin-connection
compatible with $e^a_{\mu}$.

We extend this formalism by introducing in the place of the tetrad a mixed
tensor field $V^a_\mu$ independent of the metric tensor
and the spin-connection.
The Hilbert-Palatini Lagrangian density is replaced with
\begin{equation}
\label{eq:LR}
\mathcal{L}_{R}=-\sqrt{g}\,\frac{1}{2}\,
g^{\mu\rho}g^{\nu\sigma}V^a_\mu V^b_\nu R_{ab\rho\sigma}~,
\end{equation}
where $g=|\det g_{\mu\nu}|$ and
$R_{ab\rho\sigma}=\eta_{ac}R^c_{~b\rho\sigma}$.
We introduce also a scalar field $\phi$ described by
the Lagrangian density
\begin{equation}
\label{eq:Lscalar}
\mathcal{L}_\phi=\sqrt{g}\,\left[\frac{1}{2}\,g^{\mu\nu}\partial_\mu\phi\,
\partial_\nu\phi
-V(\phi)
-\frac{\gamma}{2}\, g^{\mu\nu}\eta_{ab}V^a_\mu V^b_\nu\phi^2\right]~,
\end{equation}
where the potential is $V(\phi)=\lambda (\phi^2-v^2)^2$
and $\gamma$, $\lambda$ and $v^2$ are positive constants.
As a complement, we introduce a matter Lagrangian density $\mathcal{L}_M$
that includes a possible cosmological constant term
and, for the sake of simplicity, is independent of $\phi$ and $V^a_\mu$.

The field $V^a_\mu$ is a Lagrange multiplier that introduces a constraint
on the metric tensor, the spin-connection and the scalar field.
By varying the action with respect to $V^a_\mu$, we obtain
an eigenvalue equation similar to (\ref{eq:einstein}),
\begin{equation}
\label{eq:eigenvalue}
g^{\nu\sigma}R^{a}_{~b\mu\sigma}\,V^b_{\nu}=-\gamma\phi^2V^a_{\mu}~,
\end{equation}
which has a non-trivial solution if
\begin{equation}
\label{eq:det}
\det(g^{\nu\sigma}R^a_{~b\mu\sigma}
+\gamma\phi^2\delta^a_b\delta^{\nu}_{\mu})=0~.
\end{equation}
Therefore, $-\gamma\phi^2$ is an eigenvalue of $R^{a~~\nu}_{~b\mu}$
($=g^{\nu\sigma}R^a_{~b\mu\sigma}$) and $V^a_\mu$ is an eigenvector
belonging to it.

As we shall see below, (\ref{eq:eigenvalue}) is an extended Einstein
field equation that reduces to (\ref{eq:einstein}) in the vacuum.
Thus, the effective cosmological constant is related to the vacuum
expectation value of the scalar field by
\begin{equation}
\label{eq:cosmoconst}
\Lambda=\gamma\phi^2_{vac}~.
\end{equation}
However, the value of $\phi_{vac}$ is not $\pm v$ as set by the usual
mechanism of spontaneous symmetry-breaking,
but depends on the energy density of matter in the vacuum.
Reciprocally, the range of possible values of this energy density
is bounded by the condition that $\phi^2_{vac}\geq0$.

In order to show that (\ref{eq:eigenvalue}) is a sound extension of
the Einstein field equation, it is not enough to show that it reduces to
(\ref{eq:einstein}) in the vacuum, we also have to find out about
the connection between the curvature tensor and the energy-momentum
tensor. By varying the total action with respect to $g^{\mu\nu}$,
we obtain
\begin{equation}
\label{eq:metric}
\partial_\mu\phi\,\partial_\nu\phi
-\gamma\,\eta_{ab}V^a_\mu V^b_\nu\phi^2
-g_{\mu\nu}\left[
\frac{1}{2}\,g^{\alpha\beta}\partial_\alpha\phi\,\partial_\beta\phi
-V(\phi)
-\frac{\gamma}{2}\,g^{\alpha\beta}\eta_{ab}V^a_\alpha V^b_\beta\phi^2
\right]
+T_{R\mu\nu}+T_{M\mu\nu}=0~,
\end{equation}
where $T_{R\mu\nu}$ is a shorthand for the traceless tensor
\begin{equation}
T_{R\mu\nu}=-g^{\beta\sigma}\left(V^a_{\mu}V^b_{\beta}R_{ab\nu\sigma}
+V^a_{\nu}V^b_{\beta}R_{ab\mu\sigma}\right)
+\frac{1}{2}\,g_{\mu\nu}\,g^{\alpha\rho}g^{\beta\sigma}
V^a_{\alpha}V^b_{\beta}R_{ab\rho\sigma}
\end{equation}
and $T_{M\mu\nu}$ is the all-important energy-momentum tensor of matter,
\begin{equation}
T_{M\mu\nu}=\frac{2}{\sqrt{g}}
\frac{\delta \mathcal{L}_M}{\delta g^{\mu\nu}}~.
\end{equation}
Contracting (\ref{eq:eigenvalue}) with $\eta_{ab}V^b_{\nu}$ yields
\begin{equation}
\label{eq:eigenvalue2}
g^{\beta\sigma}V^a_{\nu}V^b_{\beta}R_{ab\mu\sigma}=
-\gamma\,\eta_{ab}V^a_{\mu}V^b_{\nu}\phi^2~.
\end{equation}
Substituting this result for the second term in (\ref{eq:metric})
and separating the marble from the wood, we obtain
\begin{equation}
\label{eq:extension}
g^{\beta\sigma}V^a_{\mu}V^b_{\beta}R_{ab\nu\sigma}
-\frac{1}{2}\,g_{\mu\nu}\,g^{\alpha\rho}g^{\beta\sigma}
V^a_{\alpha}V^b_{\beta}R_{ab\rho\sigma}=T_{\mu\nu}~,
\end{equation}
where $T_{\mu\nu}$ is the sum of the energy-momentum tensors of
the scalar field and matter,
\begin{equation}
\label{eq:totalenergy-momentum}
T_{\mu\nu}=\partial_\mu\phi\,\partial_\nu\phi
-g_{\mu\nu}\left[
\frac{1}{2}\,g^{\alpha\beta}\partial_\alpha\phi\,\partial_\beta\phi
-V(\phi)
-\frac{\gamma}{2}\,g^{\alpha\beta}\eta_{ab}V^a_\alpha V^b_\beta\phi^2
\right]+T_{M\mu\nu}~.
\end{equation}
Note that, as (\ref{eq:eigenvalue}) and (\ref{eq:metric})
together are stronger than (\ref{eq:extension}), a new field $V^a_{\mu}$
independent of $e^a_{\mu}$ is needed for mathematical consistency.
In the vacuum, however,
\begin{equation}
\label{eq:Vvacuum}
V^a_{\mu}=\frac{1}{\sqrt{8\pi G}}\,e^a_\mu
\end{equation}
and $\omega^a_{~b\mu}$ is the torsion-free spin-connection compatible with
$e^a_{\mu}$ (see below).
So far as this is a good approximation in the presence of matter,
(\ref{eq:extension}) reduces to the Einstein field equation including
matter, ``dark matter'' (the scalar field), and ``dark energy''
(the implicit cosmological constant).

By varying the action with respect to $\omega^a_{~b\mu}$, we obtain
\begin{equation}
\label{eq:spin-connection}
\partial_{\nu}\left(\sqrt{g}\,V^{\mu}_{[a}V^{\nu}_{b]}\right)
-\sqrt{g}\,\omega^c_{~a\nu}V^{\mu}_{[c}V^{\nu}_{b]}
-\sqrt{g}\,\omega^c_{~b\nu}V^{\mu}_{[a}V^{\nu}_{c]}=0~,
\end{equation}
where $V^{\mu}_{[a}V^{\nu}_{b]}=V^{\mu}_aV^{\nu}_b-V^{\mu}_bV^{\nu}_a$
and $V^{\mu}_a=\eta_{ab}\,g^{\mu\nu}\,V^b_{\nu}$.
For our present purpose, it is sufficient to observe that,
when (\ref{eq:Vvacuum}) is valid, the solution of this equation
is the unique torsion-free spin-connection compatible with $e^a_{\mu}$.

Let us now study the scalar field and the vacuum state.
First we rewrite (\ref{eq:metric}) in a more convenient form.
Contracting it with $g^{\mu\nu}$ yields
\begin{equation}
\label{eq:contracted}
-g^{\mu\nu}\partial_\mu\phi\,\partial_\nu\phi
+\gamma\,g^{\mu\nu}\eta_{ab}V^a_\mu V^b_\nu\phi^2+4V(\phi)+T_{M}=0~,
\end{equation}
where $T_{M}=g^{\mu\nu}T_{M\mu\nu}$. 
Substituting this result back into (\ref{eq:metric}), we get 
\begin{equation}
\label{eq:metricfinal}
\left[V(\phi)+\frac{1}{4}\,T_{M}\right]g_{\mu\nu}=
\partial_\mu\phi\,\partial_\nu\phi
-\gamma\,\eta_{ab}V^a_\mu V^b_\nu\phi^2+T_{R\mu\nu}+T_{M\mu\nu}
-\frac{1}{4}\,g_{\mu\nu}T_{M}~.
\end{equation}
Now we obtain the equation for the scalar field by varying the action
with respect to $\phi$,
\begin{equation}
\label{eq:scalar}
\partial_\mu\left(\sqrt{g}\,g^{\mu\nu}\partial_\nu\phi\right)
+\sqrt{g}\,\left(
\frac{dV}{d\phi}+\gamma\,g^{\mu\nu}\eta_{ab}V^a_\mu V^b_\nu\phi
\right)=0~.
\end{equation}
Multiplying this equation by $\phi$ and making use of
(\ref{eq:contracted}) to eliminate $V^a_\mu$, we get

\begin{equation}
\label{eq:phi}
\partial_\mu(\sqrt{g}\,g^{\mu\nu}\partial_\nu\phi^2)
+\sqrt{g}~8\lambda v^2(\phi^2-v^2)=\sqrt{g}~2\,T_{M}~.
\end{equation}

What we want is to find an equilibrium solution of the field equations
in which $\phi$ and all matter fields are constant in space-time.
For such constant fields, the solution of (\ref{eq:phi}) is simply
\begin{equation}
\label{eq:phivacuum}
\phi_{vac}^2=v^2+\frac{1}{4\lambda v^2}\,T_{Mvac}~,
\end{equation}
which requires that 
\begin{equation}
\label{eq:Tinequality}
T_{Mvac}\geq-4\lambda v^4~.
\end{equation}
Thus, the trace of the energy-momentum tensor of matter in the vacuum
cannot be smaller than $-4\lambda v^4$.
In other words, $\phi$ and $T_M$ are driven by spontaneous
symmetry-breaking to take vacuum expectation values such that
\begin{equation}
\lambda v^2\,\phi_{vac}^2-\frac{1}{4}\,T_{Mvac}=\lambda v^4~.
\end{equation}
However, particular values cannot be assigned to $\phi_{vac}$
and $T_{Mvac}$ without fine-tuning.

Introducing the composite field
\begin{equation}
\chi=\frac{1}{2\,|\phi_{vac}|}(\phi^2-\phi_{vac}^2)~,
\end{equation}
we can reexpress (\ref{eq:phi}) as 
\begin{equation}
\label{eq:chi}
\partial_\mu(\sqrt{g}g^{\mu\nu}\partial_\nu\chi)
+\sqrt{g}~8\lambda v^2\chi=\sqrt{g}~\frac{1}{|\phi_{vac}|}
(T_{M}-T_{Mvac})~,
\end{equation}
where there is no sign of the $\phi^4$ and $V\phi$ interactions,
the mass of $\chi$ is equal to the mass usually acquired by $\phi$
as a result of spontaneous symmetry breaking,
$m_\chi=\sqrt{8\lambda v^2}$, and the source of the field is
the trace of the energy-momentum tensor of matter above the ground state
value, even though $\phi$ is not directly coupled with matter.

Lorentz invariance requires that in the vacuum the energy-momentum tensor
takes the form
\begin{equation}
\label{eq:TMvacuum}
T_{M\mu\nu}=\rho_{Mvac}\,g_{\mu\nu}~.
\end{equation}
Consequently, (\ref{eq:metricfinal}) reduces to
\begin{equation}
\label{eq:metricvacuum}
\rho_{Mvac}\,g_{\mu\nu}=-\gamma v^2\eta_{ab}V^a_\mu V^b_\nu
\end{equation}
for $\phi_{vac}\neq0$.
This is consistent with the adopted signature of $g_{\mu\nu}$ only if
$\rho_{Mvac}$ is negative.
Taking into account (\ref{eq:Tinequality}), we get the important result
that the range of possible values of the energy density of matter
in the vacuum is restricted to the interval
\begin{equation}
\label{eq:rhoinequalities}
-\lambda v^4\leq\rho_{Mvac}\leq0~.
\end{equation}
There would be no equilibrium vacuum state if this inequality were not
satisfied.

It follows from (\ref{eq:tetrad1}) and (\ref{eq:metricvacuum}) that
in the vacuum
\begin{equation}
\label{eq:Vvacuum2}
V^a_{\mu}=\sqrt{-\frac{\rho_{Mvac}}{\gamma v^2}}\,e^a_{\mu}~.
\end{equation}
This coincides with (\ref{eq:Vvacuum}) and agrees with the Einstein field
equation if
\begin{equation}
\label{eq:rhoG}
\rho_{Mvac}=-\frac{1}{8\pi G}\,\gamma v^2~.
\end{equation}
Thus, the gravitational constant is determined by the energy density of
matter in the vacuum for a fixed value of $\gamma v^2$.
It is interesting that $G$ would be infinite if $\rho_{Mvac}$
vanished. Therefore, the Universe, which has an extravagantly feeble
gravity, cannot have a vanishingly small energy density of matter
in the vacuum.

As $\omega^a_{~b\mu}$ reduces to the usual spin-connection when
(\ref{eq:Vvacuum2}) is valid, (\ref{eq:eigenvalue2}) takes the form
of the Einstein field equation for pure gravity with a cosmological
constant,
\begin{equation}
R_{\mu\nu}=-\Lambda\,g_{\mu\nu}~,
\end{equation}
where $R_{\mu\nu}$ is the Ricci tensor and $\Lambda$ is given by
(\ref{eq:cosmoconst}).
This equation is valid not only in the vacuum, it is satisfied also
by the exterior gravitational field of any distribution of matter and
"dark matter". In particular, the successful predictions of general
relativity in the weak field regime of our solar system are recovered
if $m_{\chi}$ is not too small.
 
Taking into account (\ref{eq:phivacuum}),
(\ref{eq:TMvacuum}), and (\ref{eq:rhoG}), we get
\begin{equation}
\label{eq:LambdaG}
\Lambda=\gamma\left(v^2-\frac{1}{8\pi G}\,\frac{\gamma}{\lambda}\right)~.
\end{equation}
At the minimum allowed energy density,
$\rho_{Mvac~min}=-\lambda v^4$,
the gravitational constant also is minimum,
$8\pi G_{min}=\gamma/\lambda v^2$,
and the effective cosmological constant is zero.
Of course, this is fine-tuning.
As asserted by Weinberg's no-go theorem,
this model does not imply a vanishing cosmological constant.
What is important is that it does imply a bounded total vacuum energy
density. In fact, from (\ref{eq:rhoinequalities}), we get
\begin{equation}
\label{eq:feeble}
\gamma\leq8\pi G\lambda v^2=\pi Gm_\chi^2~.
\end{equation}
This inequality can be expressed as a restriction on the possible values
of the total vacuum energy density,
\begin{equation}
\label{eq:Lambdainequality}
0\leq\frac{\Lambda}{8\pi G}\leq\frac{\lambda v^4}{4}~.
\end{equation}
Therefore, the natural scale of the vacuum energy density is set by
the maximum of the potential energy of the scalar field, not by the
energy density of matter in the vacuum.

Inequality (\ref{eq:feeble}) means that the dimensionless coupling
$\gamma$ is as feeble as the gravitational atraction between particles
described by the composite field $\chi$.
If $m_\chi$ is smaller than the proton mass,
\begin{equation}
\gamma\leq2\times10^{-40}~.
\end{equation}
This is what we mean, quantitatively, when we say that gravity is
extravagantly feeble.\cite{Wilczek}

The fact that $\gamma$ is very small is the reason why the cosmological
constant is small in the Planck scale, independently of the actual value
of the energy density of matter in the vacuum.
To show that this is a meaningful solution of the cosmological constant
problem, it is convenient to introduce the parameter
\begin{equation}
\label{eq:alpha}
\alpha=-\frac{\rho_{Mvac}}{\lambda v^4}~,
\end{equation}
which according to (\ref{eq:rhoinequalities}) can take any value
in the interval $0\leq\alpha\leq1$.
If we choose the unity of energy such that $\lambda v^4=1$,
$\alpha$ is equal to the absolute value of the energy density of matter
in the vacuum, the gravitational constant is
\begin{equation}
\label{eq:Galpha}
8\pi G=\frac{1}{\alpha}\,\frac{\gamma}{\sqrt{\lambda}}~,
\end{equation}
and the cosmological constant is
\begin{equation}
\label{eq:Lambdaalpha}
\Lambda=(1-\alpha)\frac{\gamma}{\sqrt{\lambda}}~.
\end{equation}
Hence the total vacuum energy density is given by
\begin{equation}
\label{eq:LambdaG2}
\frac{\Lambda}{8\pi G}=\alpha(1-\alpha)
\end{equation}
and the cosmological constant in the Planck scale is given by
\begin{equation}
\label{eq:GLambda2}
8\pi G\,\Lambda=\frac{1-\alpha}{\alpha}\,\frac{\gamma^2}{\lambda}~.
\end{equation}
At $\alpha=1/2$, the total vacuum energy density is maximum,
\begin{equation}
\left[\frac{\Lambda}{8\pi G}\right]_{max}=\frac{1}{4}~,
\end{equation}
and the cosmological constant in the Planck scale is
\begin{equation}
8\pi G\,\Lambda=\frac{\gamma^2}{\lambda}~.
\end{equation}

Astronomical observations indicate that $8\pi G\,\Lambda\approx 10^{-120}$
and $\lambda v^4\approx1~\text{meV}^4$.
As there are good reasons to believe that neither $\alpha\approx0$
(no fine-tuning without the scalar field) nor $\alpha\approx1$
(no fine-tuning with the scalar field), we are naturally led to assume
\begin{equation}
\label{eq:tiny}
\frac{\gamma}{\sqrt{\lambda}}\approx 10^{-60}~.
\end{equation}
According to (\ref{eq:Galpha}) and (\ref{eq:Lambdaalpha}),
the gravitational constant is very small for $10^{-60}\ll\alpha\leq1$ and
the effective cosmological constant is very small for all values
of $\alpha$. No fine-tuning is needed, the energy density of matter can
take any value in the interval (\ref{eq:rhoinequalities}).

In conclusion, the properties of the equilibrium vacuum state enforce
the cancellation of all the large and apparently disparate contributions
to the vacuum energy density estimated by the quantum field theory,
to produce a net value consistent with
(\ref{eq:rhoinequalities}) and (\ref{eq:Lambdainequality}). 
No anthropic considerations are needed, but the tiny value of
(\ref{eq:tiny}) remains a deep mystery.


\begin{thebibliography}{99}
\bibitem{Weinberg89} S. Weinberg, Rev. Mod. Phys. 61, 1 (1989).
\bibitem{Weinberg96} S. Weinberg, arXiv:astro-ph/9610044.
\bibitem{Rugh} S.E. Rugh and H. Zinkernagel, arXiv:hep-th/0012253.
\bibitem{Carroll} S.M. Carroll, Living Rev. Relativity \textbf{4},
1 (2001), http://www.livingreviews.org/Articles/Volume4/2001-1carroll/.
\bibitem{Garriga} J. Garriga and A. Vilenkin, Phys. Rev. D 64 (2001)
023517, arXiv:hep-th/0011262.
\bibitem{Straumann} N. Straumann, arXiv:astro-ph/0203330.
\bibitem{Peebles} P.J.E. Peebles and B. Ratra, arXiv:astro-ph/0207347.
\bibitem{Pad} T. Padmanabhan, arXiv:hep-th/0212290.
\bibitem{Dolgov} A.D. Dolgov, Phys. Rev. D 55, 5881 (1997).
\bibitem{Peldan} P. Peld\'an, arXiv:gr-qc/9305011.
\bibitem{Wilczek} F. Wilczek, arXiv:hep-ph/0201222.
\end{thebibliography}
\end{document}